\documentclass[11pt,twoside]{article}

\usepackage{asp2004}
\usepackage{epsf}
\usepackage{psfig}
\usepackage{lscape}

\begin{document}
\title{Dust and Gas Debris Around Main Sequence Stars}   %%% Fill in title
\author{Christine H. Chen}   %%% Fill in author names
\affil{National Optical Astronomy Observatory; P.O. Box 26732; 950 N. Cherry Ave.; Tucson, AZ 85726-6732}    %%% Fill in author affiliations

\begin{abstract} %%% Abstract to run on from here.
Debris disks are dusty, gas-poor disks around main sequence stars (Backman \&
Paresce 1993; Lagrange, Backman \& Artymowicz 2000; Zuckerman 2001).
Micron-sized dust grains are inferred to exist in these systems from 
measurements of their thermal emission at infrared through millimeter 
wavelengths. The estimated lifetimes for circumstellar dust grains 
due to sublimation, radiation and corpuscular stellar wind effects are 
typically significantly smaller than the estimated ages for the stellar 
systems, suggesting that the grains are replenished from a reservoir, such as 
sublimation of comets or collisions between parent bodies. Since the color 
temperature for the excess emission is typically $T_{gr}$ $\sim$ 110 - 120 K, 
similar to that expected for small grains in the Kuiper Belt, these objects are
believe to be generated by collisions between parent bodies analogous to Kuiper
Belt objects in our solar system; however, a handful of systems possess warm 
dust, with $T_{gr}$ $\geq$ 300 K, at temperatures similar to the terrestrial 
planets. We describe the physical characteristics of debris disks, the 
processes that remove dust from disks, and the evidence for the presence of 
planets in debris disks. We also summarize observations of infalling comets 
toward $\beta$ Pictoris and measurements of bulk gas in debris disks.
\end{abstract}

%%% MAIN BODY OF TEXT GOES HERE. CONSULT "INSTRUCTIONS FOR AUTHORS USING
%%% LATEX2E MARKUP", SECTIONS 2.3-2.6 FOR HELP WITH EQUATIONS, FIGURES,
%%% AND TABLES.

\section{Introduction}
The observation of periodic variations in the radial velocities of nearby
late-type stars has led to the discovery of $\sim$150 giant planets (Marcy,
Cochran, \& Mayor 2000), including 14 multiple planet systems, suggesting that
planetary systems may exist around $>$5\% of solar-like main sequence stars. 
Giant planets, like Jupiter, are believed to form in proto-planetary disks made
of gas and dust. As grains grow into larger bodies, the bulk of circumstellar 
gas dissipates from the system (e.g., via accretion onto the star or planetary 
objects) until the disk contains predominately large bodies, such as planets 
and/or planetesimals and dusty debris. Studying the dynamics of dust and gas in
disks around main sequence stars reveals the presence of planets and small 
bodies, analogous to asteroids and comets in our solar system. Therefore, 
studying debris disks may allow us to determine the processes that shape the 
final architectures of planetary systems including our own solar system.

The lifetimes of dust grains in debris disks are believed to be significantly 
shorter than the ages of the systems, suggesting that the particles are 
replenished from a reservoir such as collisions between parent bodies. Our 
solar system shows evidence for collisions within the main asteroid belt that
produce micron-sized dust grains. In 1918, Hirayama discovered concentrations 
of asteroids in three regions of $a-e-i$ ($a$ is the osculatory orbital 
semi-major axis; $e$ is the eccentricity, and $i$ is the inclination) space 
which he named the Themis, Eos, and Koronis families. The clumping of these 
asteroids is widely attributed to the break up of larger parent bodies (Chapman
et al. 1989). \emph{IRAS} observations of the zodiacal dust discovered the 
$\alpha$, $\beta$, and $\gamma$ dust bands which have orbital properties 
identical to the Themis, Koronis, and Eos families, suggesting that the dust in
each band was created by collisions between asteroids in each family. 
Non-equilibrium processes may be responsible for the generation of dust bands. 
Gravitational perturbations by Jupiter and other planets in our solar system 
are expected to cause the apsides and nodes of asteroid orbits to precess at 
different rates because of small differences in their orbital parameters. This 
precession leads to asteroid collisions that generate the small grains observed
in the dust bands.

Observed debris disks typically possess luminosities 3 - 5 orders of magnitude 
higher than our zodiacal disk, suggesting that they are significantly more 
massive. The dust in these systems may be produced in steady state at earlier 
times when the planetesimal belt was more massive or during an epoch of intense
collisions that produced especially large quantities of dust, triggered by the 
formation and/or migration of planets. Simulations of planetesimal disks,
in which planets are forming from pairwise collisions, suggest that formation
of massive planets may trigger collisional cascades between the remaining 
nearby planetesimals (Kenyon \& Bromley 2004). Alternately, the migration of 
giant planets, soon after their formation, may also trigger collisions. In our 
solar system, the moon and terrestrial planets experienced an intense period of
cratering 3.85 Gyr ago known as the Period of Late Heavy Bombardment. Two 
pieces of evidence suggest that the impactors during this period may have been 
asteroids. (1) Lunar impact melts, collected during Apollo missions, suggest 
that the composition of impactors is similar to asteroids. (2) The size 
distribution of impactors, inferred from the lunar highlands, is similar to
that of the main asteroid belt. The migration of the Jovian planets during
the Late Heavy Bombardment may have caused gravitational resonances to sweep 
through the main asteroid belt sending asteroids into the inner solar system 
(Strom et al. 2005). 

\section{Gross Properties}
The debris disks around Vega, Fomalhaut, $\epsilon$ Eridani, and $\beta$
Pictoris were initially discovered from the presence of strong \emph{IRAS}
60 $\mu$m and 100 $\mu$m fluxes, 10 - 100 times larger than expected from the
photosphere alone (Backman \& Paresce 1993). Studies comparing the \emph{IRAS}
fluxes with predictions for the photospheric emission of field stars
subsequently discovered more than 100 debris disk candidates (e.g., Walker \& 
Wolstencroft 1988; Oudmaijer et al. 1992; Backman \& Paresce et al. 1993; 
Sylvester et al. 1996; Mannings \& Barlow 1998; Silverstone 2000). The launch 
of the \emph{Spitzer Space Telescope} (Werner et al. 2004), with unprecedented 
sensitivity in the far-infrared, is making the discovery and statistical study 
of large numbers of debris disks possible. The first exciting results from this
mission are determining the mean properties of debris disks and measuring the 
characteristic timescales for dust and gas decay. \emph{IRAS} and 
\emph{Spitzer} MIPS surveys of main sequence stars now suggest that 
$\sim$10 - 15\% of A- through K-type field stars possess 10 $\mu$m - 200 $\mu$m
excesses indicative of the presence of circumstellar dust (Bryden et al. 2005; 
Decin et al. 2003). However, surveys of M-type stars suggest that the 
fraction of dwarfs with infrared excesses is lower (Gautier et al. 2004; Song 
et al. 2002); only 2 out of 150 \emph{IRAS}/\emph{Spitzer} detected M-dwarfs 
possess far-infrared excess: Hen 3-600 in the TW Hydrae Association and AU Mic 
in the $\beta$ Pic Moving Group, with estimated ages $\sim$10 Myr. 

{\bf Fractional Infrared Luminosity:} The average fraction of stellar 
luminosity reprocessed by the circumstellar dust grains in debris disks 
discovered by \emph{IRAS} and \emph{ISO}, $L_{IR}/L_{*}$, is typically between 
$\sim$10$^{-5}$ and $\sim$5$\times$10$^{-3}$ (Decin et al. 2003). As a result, 
the majority of debris disks discovered using these satellites are located 
around high luminosity A-type stars. The improved sensitivity of \emph{Spitzer}
may drive the discovery of debris disks at 70 $\mu$m and 160 $\mu$m with 
$L_{IR}/L_{*}$ as faint as 10$^{-8}$. For comparison, the zodiacal dust in our 
solar system reprocesses, $L_{IR}/L_{*}$ = $10^{-7}$ (Backman \& Paresce 1993),
and the dust in the Kuiper Belt is estimated to reprocess, $L_{IR}/L_{*}$ = 
$10^{-7}$ - $10^{-6}$ (Backman, Dasgupta, \& Stencel 1995).

{\bf Grain Size:} The similarity between black body grain distances, inferred 
from SED modeling of \emph{IRAS} fluxes, with the measured radii of debris 
disks in resolved systems implies that the circumstellar dust grains in debris 
disks are large. For example, a black body fit to the \emph{IRAS} HR 4796A 
excess SED implies an estimated dust grain temperature, $T_{gr}$ = 110 K, 
corresponding to a distance of 35 AU if the grains are large (the absorption 
coefficient, $Q_{abs}$ $\propto$ $(2 \pi a/\lambda)^{0}$; Jura et al. 1998). 
However, if the grains are small ($2 \pi a$ $<$ $\lambda$), then they radiate 
less efficiently ($Q_{abs}$ $\propto$ $(2 \pi a/\lambda)$) and are expected to 
be located at a distance of 280 AU from the star. High resolution mid-infrared 
and coronagraphic imaging has resolved a narrow ring of dust around HR 4796A at
a distance of 70 AU from the star, more consistent with large grains (Schneider
et al. 1999; Telesco et al. 2000). In addition, \emph{Spitzer} IRS spectroscopy
(at 5 $\mu$m - 35 $\mu$m) of 60 main sequence stars with \emph{IRAS} 60 $\mu$m 
excesses (including HR 4796A) do not detect silicate emission feature at 
10 $\mu$m and 20$\mu$m, suggesting that the grains have radii, $a$ $>$ 10 
$\mu$m (Jura et al. 2004; Chen et al. 2006, in preparation). 

{\bf Disk Mass:} Scattered light and far-infrared imaging observations are
sensitive to dust grains with radii $<$100 $\mu$m; however, the majority of the
disk mass is probably contained in larger objects that do not possess as much 
surface area. Therefore, measurements of disk mass are made at submillimeter
wavelengths where the grains are more likely to be optically thin. The typical 
dust mass in debris disks, is 0.01 $M_{\earth}$ - 0.25 $M_{\earth}$ (Zuckerman
\& Becklin 1993; Holland et al. 1998; Najita \& Williams), a tiny fraction of 
the 10 - 300 $M_{\earth}$ measured toward pre-main sequence T-Tauri and Herbig 
Ae/Be stars (Natta, Grinin, \& Mannings 2000) and a tiny fraction of the dust 
mass expected in a minimum mass solar nebula with an interstellar gas:dust 
ratio ($\sim$30 $M_{\earth}$). The decline in submillimeter flux, as objects 
age from Herbig Ae stars to main sequence stars, may be the result of accretion
of grains onto the central star or grain growth into larger bodies. After stars
reach the main sequence, the decline in submillimeter flux may be the result of
parent body grinding. The upper envelope of disk masses for A-type stars is 
higher than that of later-type stars, suggesting that high mass stars may 
initally possess more massive disks (Liu et al. 2004a). For comparison, the 
estimated total mass in the main asteroid belt is $\sim$0.0003 $M_{\earth}$; 
the asteroid belt may have contained as much as 1000$\times$ more mass prior to
the period of Late Heavy Bombardment. The estimated total mass in the Kuiper 
Belt is 0.1 $M_{\earth}$.

{\bf Gas:Dust Ratio} Current (uncertain) models suggest that Jovian planets 
form either via rapid gravitational collapse through disk instability within a 
few hundred years (Boss 2003) or via coagulation of dust into solid cores 
within the first $\sim$1 Myr and accretion of gas into thick hydrogen 
atmospheres within the first $\sim$30 Myr (Pollack et al. 1996). At present, 
the timescales on which giant planets form and accrete their atmospheres have 
not been well constrained observationally. CO surveys suggest that the bulk of 
molecular gas dissipates within the first $\sim$10 Myr (Zuckerman, Forveille, 
\& Kastner 1995). Since the bulk disk gas is expected to be composed largely of
H$_{2}$, recent surveys have focused on searching for this molecule. New 
high-resolution (R = 600) \emph{Spitzer} IRS observations place upper limits on
the H$_{2}$ S(0) and S(1) emission toward $\beta$ Pictoris, 49 Ceti, and 
HD 105 that suggest that $<$1-15 $M_{\earth}$ H$_{2}$ remains in these systems
(Chen et al. 2006; Hollenbach et al. 2005). The gas:dust ratio in debris disks 
is probably $<$10:1. UV absorption line studies have placed upper limits on the
gas:dust ratios two nearly edge-on systems. \emph{FUSE} and \emph{SITS} 
observations constrain the gas:dust ratio in the AU Mic disk ($<$6:1) by 
placing upper limits on the H$_{2}$ absorption in the O VI 
$\lambda \lambda$1032, 1038 emission lines and in the C II 
$\lambda \lambda$1036, 1037 and $\lambda$1335 emission lines (Roberge et al. 
2005). \emph{FUSE} and \emph{SITS} observations constrain the gas:dust ratio in
the HR 4796A disk ($<$4:1) by placing upper limits on H$_{2}$ and 
$\lambda$2026 Zn II  absorption (Chen \& Kamp 2004).

\section{Debris Disks as Solar System Analogs}
Parallels are often drawn between parent body belts in debris disks and small 
body belts in our solar system. These analogies are based on the similarity of 
observed grain temperatures to those inferred for the main asteroid belt and 
the Kuiper Belt.

\subsection{Dust in Extra-Solar Kuiper Belts}
Numerous small bodies have been discovered in our solar system at distances
of 30 - 50 AU from the Sun. These bodies are collectively referred to as the
Kuiper Belt and are the likely source for short-period comets in our solar 
system. Kuiper Belt objects are expected to collide and grind down into dust
grains that may be detected at far-infrared wavelengths (Backman, Dasgupta, \&
Stencel 1995). If these dust grains are large, then they are expected to have 
grain temperatures, $T_{gr}$ = 40 - 50 K; if they are small, then $T_{gr}$ 
= 110 - 130 K. The bulk of the energy radiated by grains with these 
temperatures should emerge at 30 $\mu$m to 90 $\mu$m, consistent with the 
properties of debris disks discovered using \emph{IRAS} (Backman \& Paresce 
1993); therefore, the majority of discovered debris disks are envisioned to be 
massive analogs to the Kuiper Belt. \emph{Spitzer} IRS spectroscopy of 60 
\emph{IRAS}-discovered debris disks suggests that the 5 $\mu$m - 35 $\mu$m 
spectra of \emph{IRAS} 60 $\mu$m excess sources can be modeled using a single 
temperature black body (Chen et al. 2006; Jura et al. 2004). The peak of the 
IRS inferred grain temperature distribution lies at $T_{gr}$ = 110 - 120 K; 
although, the lack of data beyond 35 $\mu$m makes this study insensitive to 
grains with $T_{gr}$ $<$65 K. \emph{Spitzer} MIPS surveys of main sequence FGK 
stars have discovered a number of solar-like stars with 70 $\mu$m excess but no
24 $\mu$m excess, consistent with $T_{gr}$ $<$ 100 K (Bryden et al. 2005; Chen 
et al. 2005b; Kim et al. 2005).  

The low luminosity of M-type stars makes detecting thermal emission from
circumstellar dust around these stars challenging compared to detecting 
thermal emission from dust around high luminosity A-type stars. For example, 
dust located 50 AU from an M2V star is expected to possess $T_{gr}$ = 17 K if 
$L_{*}$ = 0.034 $L_{\sun}$ and the grains are black bodies. Since the dust 
around M-type stars is so cool, disks around M-dwarfs may be detected best at 
submillimeter wavelengths. At the distance of the closest stars (50 pc), a
disk that reprocesses $L_{IR}/L_{*}$ = 2$\times$10$^{-3}$ would produce an 
undetectable flux at 100 $\mu$m, F$_{\nu}$(100 $\mu$m) $<$10 mJy. However, it 
would produce a robust excess at submillimeter wavelengths, 
F$_{\nu}$(850 $\mu$m) = 30 mJy. One star has been detected at submillimeter 
wavelengths depite the lack of \emph{IRAS} excess. JCMT SCUBA photometry of 
TWA 7 at 850 $\mu$m detects thermal emission from dust, with 
$F_{\nu}$(850 $\mu$m) = 15.5 mJy (Webb 2000), even though the source does not 
appear in the \emph{IRAS} catalog.  A survey of three young M-dwarfs in 
the $\beta$ Pic Moving Group (with an estimated age of 12 Myr) and the Local 
Association Group (with an estimated age of 50 Myr) discovered 450 $\mu$m
and/or 850 $\mu$m excesses associated with two of the stars: AU Mic and Gl 182 
(Liu et al. 2004a). The disk around AU Mic is warm enough and massive enough 
that its disk is bright at 70 $\mu$m (F$_{\nu}$(70 $\mu$m) = 200 mJy)
while the disk around Gl 182 is not detected at either 24 $\mu$m or 70 $\mu$m
(Chen et al. 2005b).

\subsection{Dust in Extra-Solar Asteroid Belts}
The zodiacal dust in our solar system possesses a grain temperature, $T_{gr}$ =
150 - 170 K, suggesting that the bulk of the thermal energy is radiated at
20 $\mu$m - 25 $\mu$m. The Earth has a temperture, $T_{gr}$ $\sim$ 300 K, 
suggesting that the bulk of its thermal energy is radiated at $\sim$10 $\mu$m. 
Searches for 10 $\mu$m excesses around main-sequence stars have revealed that 
300 K dust around A - M dwarfs is rare. \emph{IRAS} surveys of main sequence
stars suggest that $<$5\% of debris disks possess 12 $\mu$m excesses. In a 
survey of 548 A - K dwarfs, Aumann \& Probst (1991) were able to identify 
\emph{IRAS} 12 $\mu$m excesses only with $\beta$ Pictoris and $\zeta$ Lep.
Identifying objects with 12 $\mu$m excess is challenging because the 
photosphere usually dominates the total flux at this wavelength. High 
resolution mid-infrared imaging using LWS at Keck confirms that the debris disk
around $\zeta$ Lep is compact. The disk is at most marginally resolved at 
17.9 $\mu$m, suggesting that the dust probably lies within 6 AU althought some 
dust may extend as far as 9 AU away from the star, consistent with the 230 K 
- 320 K color temperature inferred 10 $\mu$m - 60 $\mu$m photometry (Chen \& 
Jura 2001). Similarly, 12 $\mu$m excess around M-dwarfs appears to be rare. 
Follow-up Keck LWS 11.7 $\mu$m photometry of nine late-type dwarfs with 
possible \emph{IRAS} 12 $\mu$m excesses is unable to detect excess thermal 
emission from any of the candidates (Plavchan et al. 2005).

Warm silicates with grain temperatures, $T_{gr}$ $\sim$300 K, may produce
emission in the 10$\mu$m Si-O stretch mode, if the dust grains are small, 
$a$ $<$10 $\mu$m. These features can provide insight into not only the
composition but also the size of circumstellar dust grains. More than a decade 
ago, ground-based spectroscopy of $\beta$ Pictoris revealed a broad 9.6 $\mu$m 
amorphous silicate and a weaker 11.2 $\mu$m crystalline olivine emission 
feature, similar to that observed toward comets Halley, Bradford, and Levy 
(Knacke et al. 1993), suggesting that the parent bodies may be similar to small
bodies in our solar system. Several systems with T$_{gr}$ $>$300 K have now 
been discovered that possess silicate emission features. \emph{Spitzer} IRS 
spectra of HD 69830, a K0V star with an age of 2 Gyr, show mid-infrared 
emission features nearly identical to those observed toward Hale-Bopp but with 
a higher grain temperature, $T_{gr}$ = 400 K instead of $T_{gr}$ = 207 K 
(Beichman et al. 2005b). Gemini Michelle spectroscopy of BD+20 307 (HIP 8920), 
a G0V star with an estimated age of 300 Myr, suggests that amorphous and 
crystalline silicates are present; models of the 9 - 25 $\mu$m SED imply 
$T_{gr}$ = 650 K and a remarkly high $L_{IR}/L_{*}$ = 0.04 for its age (Song et
al. 2005). More warm dust systems with spectral features may soon be identified
using \emph{Spitzer} MIPS. For example, $\alpha^1$ Lib (a F3V star in the 
200 Myr old Castor Moving Group) and HD 177724 (an A0V field star) possess such
strong 24 $\mu$m excess that their 12 $\mu$m, 24 $\mu$m, and 70 $\mu$m fluxes 
can not be self-consistently modeled using a modified black body, suggesting 
that their strong 24 $\mu$m excessess may be the result of emission in spectral
features (Chen et al. 2005b).

Since SED modeling is degenerate, high resolution imaging is needed to 
determine definitely the location of the dust and to search for structure
in the disk. Nulling interferometers, now operational at Keck and the 
Multiple-Mirror Telescope (MMT), will allow dust in exo-zodical disks to be 
directly resolved. By placing the central star in a null, nulling 
interferometry destructively interfers stellar emission, obviating the need for
accurate models of the stellar atmosphere. Without the bright central core, 
nulling observations can not only seek faint exo-zodial emission but can also 
apply the full diffraction limit of the telescope to resolve a source. For 
example, high-resolution Keck LWS imaging of the Herbig Ae star AB Aur at 
18.7 $\mu$m (with a FWHM 0.5$\arcsec$) struggles to resolve faint extended 
emission in the wings of the PSF at 0.4$\arcsec$ from the star (Chen et al. 
2003) while MMT nulling observations at 10.6 $\mu$m suppress all but 10\% 
- 20\% of the flux; fits of the percentage null versus rotation of the 
interferometer baseline suggest that the mid-infrared emitting component 
possess an angular diameter $\sim$0.2$\arcsec$ with a position angle, 
30$^{\circ} \pm 15^{\circ}$, and an inclination, 45$^{\circ}$ - 65$^{\circ}$, 
from face-on (Liu et al. 2005). Nulling observations of Vega at 10.6 $\mu$m do 
not detect resolved emission at $>$2.1\% (3$\sigma$ limit) of the level of the 
stellar photospheric emission, suggesting that Vega possess $<$650$\times$ as 
much zodiacal emission as our solar system (Liu et al. 2004b).

\section{Dust Removal Processes}
Infrared spectroscopy and SED modeling suggest that the majority of debris 
disks possess central clearings. They may be generated by planets that 
gravitationally eject dust grains that are otherwise spiralling toward their
orbit centers under Poynting-Robertson and stellar wind drag. However, a number
of other processes may also contribute to presence of absence of central 
clearings:

{\bf Sublimation:} If the grains are icy, then ice sublimation may play an 
important role in the destruction of grains near the star and may provide a 
natural explanation for the presence of central clearings in debris disks
(Jura et al. 1998) implied from black body fits to \emph{Spitzer} IRS spectra
(Chen et al. 2006). The peak in the measured $T_{gr}$ distribution, estimated 
from black body fits to \emph{Spitzer} IRS spectra, suggest that grains in 
debris disks typically have $T_{gr}$ = 110 - 120 K (Chen et al. 2006), near the
sublimation temperature of water ice in a vacuum, $T_{sub}$ = 150 K. Since 
sublimation lifetimes are sensitively dependent on grain temperature, cool 
grains may possess ices. For example, 3.5 $\mu$m grains with $T_{gr}$ = 70 K, 
have a sublimation lifetime, $T_{subl}$ = 1.3 $\times$ 10$^{7}$ Gyr while 
16 $\mu$m grains with $T_{gr}$ = 160 K, have a sublimation lifetime, 
$T_{subl}$ = 7.4 minutes! Since the majority of debris disks have $T_{gr}$ $<$ 
120 K, they could possess icy grains that may be detectable with 
\emph{Spitzer}. Crystalline water ice possesses an emission feature at 
61 $\mu$m. Low resolution (R = $\lambda$/$\Delta \lambda$ = 15 - 25) 
\emph{Spitzer} MIPS SED mode observations may be able to detect emission from 
ices in debris disks.

{\bf Radiation Effects:} The initial discovery of debris disks around high
luminosity main sequence B- and A-type stars led to speculation that radiation
pressure and the Poynting-Robertson effect may govern grain dynamics. The 
force due to radiation pressure acting on grains around A-type stars with radii
$<$few $\mu$m is larger than the force due to gravity; therefore, small grains 
are expected to be effectively removed from the circumstellar environment on 
timescales $\sim$10$^{4}$ years (Artymowicz 1988). Larger particles are subject
to the Poynting-Robertson effect in which dust grains lose angular momentum
through interactions with outflowing stellar photons. As a result, larger
grains spiral in toward their orbit center on timescales typically $<$1 Myr
(Burns, Lamy, \& Soter 1979). If a debris disk is composed of large black
body grains which spiral inward under PR drag, then the dust should be 
contained in a continuous disk with constant surface density and an infrared
spectrum, $F_{\nu}$ $\propto$ $\lambda$, at short wavelengths (Jura et al.
1998; Jura et al. 2004). \emph{Spitzer} IRS spectroscopy of 
\emph{IRAS}-discovered debris disks may have revealed one object that possess
an excess spectrum better fit by $F_{\nu}$ $\propto$ $\lambda$ than a black
body (HR 6670; Chen et al. 2006). Wyatt (2005) estimates that the grain density
in \emph{IRAS}-discovered debris disks is more than an order of magnitude too 
high for grains to migrate inward under Poynting-Robertson drag without 
suffering destructive collisions.

{\bf Corpuscular Stellar Wind Effects:} The recent discovery of debris disks 
around solar-like and M-type stars has led to speculation that corpuscular 
stellar winds may contribute to grain removal in a manner analogous to 
radiation pressure and the Poynting-Robertson effect (Plavchan, Jura, \& Lipscy
2005): (1) An outflowing corpuscular stellar wind produces a pressure on dust 
grains which overcomes the force due to gravity for small grains; however, the 
corpuscular stellar wind in only important for low luminosity stars (M-dwarfs) 
with strong stellar winds ($\sim$100 ${\dot M}_{\sun}$; Chen et al. 2005b). 
(2) Large particles orbiting the star are subject to a drag force produced when
dust grains collide with protons in the stellar wind. These collisions decrease
the velocities of orbiting dust grains and therefore their angular momentum, 
causing them to spiral in toward their orbit center. Stellar wind drag may 
explain the observed anti-correlation between \emph{Spitzer} 24 $\mu$m excess 
and \emph{ROSAT} fluxes toward F-type stars in the 3 - 20 Myr Sco-Cen (Chen et 
al. 2005a) and the lack of 12 $\mu$m excesses observed toward nearby, 
$>$10 Myr-old, late-type M-dwarfs (Plavchan et al. 2005). Recently, Strubbe 
\& Chiang (2005) have reproduced the radial brightness profile of the AU Mic 
disk assuming that dust grains, produced in collisions between parent bodies 
on circular orbits at 43 AU, generate the resolved scattered light. In their 
model, large grains produce a surface density, $\sigma$ $\propto$ $r^{0}$, at 
r $<$ 43 AU, under corpuscular and Poynting-Robertson drag modified by 
collisions; while, small grains, that are barely bound under corpuscular 
stellar wind and radiation pressure, produce a surface density, $\sigma$ 
$\propto$ $r^{-5/2}$, in the outer disk.

{\bf Collisions:} If the particle density within the disk is high, then 
collisions may shatter larger grains into smaller grains that may be removed by
radiation pressure and/or corpuscular stellar wind pressure. \emph{Spitzer} 
MIPS imaging has recently resolved symmetric extended emission from the face-on
disk around Vega with radii $>$330 AU, $>$540 AU, and $>$810 AU at 24 $\mu$m, 
70 $\mu$m, and 160 $\mu$m, respectively, that may be explained by small grains 
that are radiatively driven from the system (Su et al. 2005). Comparison of 
the 24 $\mu$m emission with the 70 $\mu$m emission suggests that the system 
possesses 2 $\mu$m grains, well below the blow-out limit of 14 $\mu$m. 
Statistical studies of the decline in fractional infrared luminosity may as a 
function of time also shed light on the processes by which dust grains are 
removed. The fractional infrared luminosity of a debris disk is expected to 
decrease inversely with time, $L_{IR}/L_{*}$ $\propto$ 1/$t_{age}$, if 
collisions are the dominant grain removal process and is expected to decrease 
inversely with time squared, $L_{IR}/L_{*}$ $\propto$ 1/$t_{age}^{2}$, if 
corpuslcular stellar wind and Poynting-Robertson drag are the dominant grain 
removal processes (Domink \& Decin 2003). \emph{Spitzer} MIPS, IRS and 
submillimeter studies of thermal emission from debris disks are consistent with
a $1/t$ decay and a characteristic timescale, $t_{o}$ $\sim$100 - 200 Myr 
(Chen et al. 2006; Najita \& Williams 2005; Liu et al. 2004a). 

\begin{figure}
\plotone{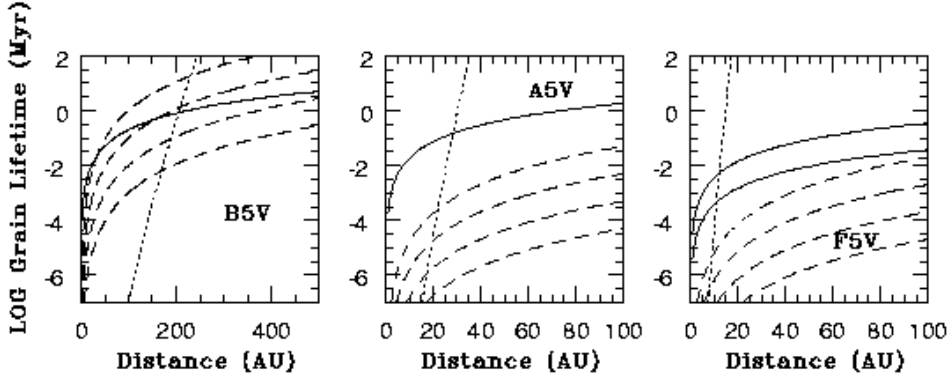}
\caption{(a) The grain lifetimes are plotted as a function of age around a B5V 
star. The Poynting-Robertson Drag/Stellar Wind drag lifetime is shown with a 
solid line; the sublimation lifetime is shown with a dotted line; and the 
collisional lifetime is shown with a dashed line, assuming $M_{dust}$ = 0.001, 
0.01, 0.1, and 1 $M_{\earth}$ (from top to bottom). (b) same as (a) for an A5V 
star. No stellar wind drag is assumed. (c) same as (a) for a F5V star. The 
corpuscular stellar wind drag lifetime is shown with a solid line, assuming 
that ${\dot M_{wind}}$ = 100, 1000 ${\dot M_{\sun}}$ (from top to bottom).}
\end{figure}

The dominant grain removal process within a disk is dependent not only on
the luminosity of the central star and the strength of its stellar wind but 
also on grain distance from the central star. For example, Backman \& 
Paresce (1993) estimate that collisions to small grain sizes and radiation 
pressure remove grains at 67 AU around Fomalhaut while the Poynting-Robertson 
effect removes grains at 1000 AU. In Figure 1, we plot the sublimation 
lifetime, the Poynting-Robertson (and corpuscular stellar wind) drag lifetime, 
and the collision lifetime for average-sized grains around typical B5V, A5V, 
and F5V stars. Sublimation may quickly remove icy grains in the innermost 
portions of the disk. At larger radii, collisions dominate grain destruction, 
and at the largest radii, where the disk has the lowest density, 
Poynting-Robertson and corpuscular stellar wind drag may dominate grain 
destruction. For typical A5V and F5V stars, the collision lifetime is shorter 
than the drag lifetime if the disk has a dust mass between 0.001 $M_{\earth}$ 
and 1 $M_{\earth}$, even if the F5V star has a stellar wind with a mass loss 
rate as high as $\dot{M}_{wind}$ = 1000 $\dot{M}_{\sun}$. However, for a 
typical B5V star, the Poynting-Robertson and stellar wind drag lifetime may be 
shorter than the collision lifetime, especially at large radii, if the disk has
a dust mass, $M_{tot}$ $<$ 0.1 $M_{\earth}$.

{\bf Gas-Grain Interactions:} In disks with gas:dust ratios between 0.1 and 10,
gas-grain interactions are expected to concentrate the smallest grains in the 
disk, with radii just above the blow-out size, at the outer edge of the disk, 
creating a ring of bright thermal emission. The presence of gas has been used 
to explain the central clearing in the HR 4796A disk (Takeuchi \& Artymowicz 
2001). 

\section{The Planet/Debris Disk Connection}
The detection of asymmetries in the azimuthal brightness of debris disks can
distinguish between whether the central clearings are dynamically sculpted by
a companion or generated by other mechanisms. For example, a planet may create 
brightness peaks in a disk by trapping grains into mean motion resonances (Liou
\& Zook 1999; Quillen \& Thorndike 2002). High resolution submillimeter imaging
of $\epsilon$ Eri has revealed the presence of brightness peaks, that may be 
explained by dust trapped into the 5:3 and 3:2 exterior mean motion resonances 
of a 30 $M_{\earth}$ planet, with eccentricity, $e$ = 0.3, and semimajor axis, 
$a$ = 40 AU (Ozernoy et al. 2000; Quillen \& Thorndike 2002). Comparison of the
1997-1998 850 $\mu$m SCUBA map with the 2000-2002 850 $\mu$m SCUBA map, in 
Figure 2, indicates that three of the brightness peaks in the ring around 
$\epsilon$ Eri are orbiting counterclockwise at a rate of 1$^{\circ}$ 
yr$^{-1}$, consistent with that expected from planetary resonance models 
(Greaves et al. 2005; Ozernoy et al. 2000). The presence of two brightness 
peaks in submillimeter and millimeter maps of the face-on disk around Vega 
(Wilner et al. 2002) may also be explained by dust trapped, this time, into the
3:2 and 2:1 exterior mean motion resonances of a Neptune-mass planet that 
migrated from 40 AU to 65 AU over a period of 56 Myr (Wyatt 2003). However, the
observation of orbital motion, the detection of the putative planet, and the 
observation of lower-level brightness asymmetries are needed to confirm this 
model.

\begin{figure}
\plottwo{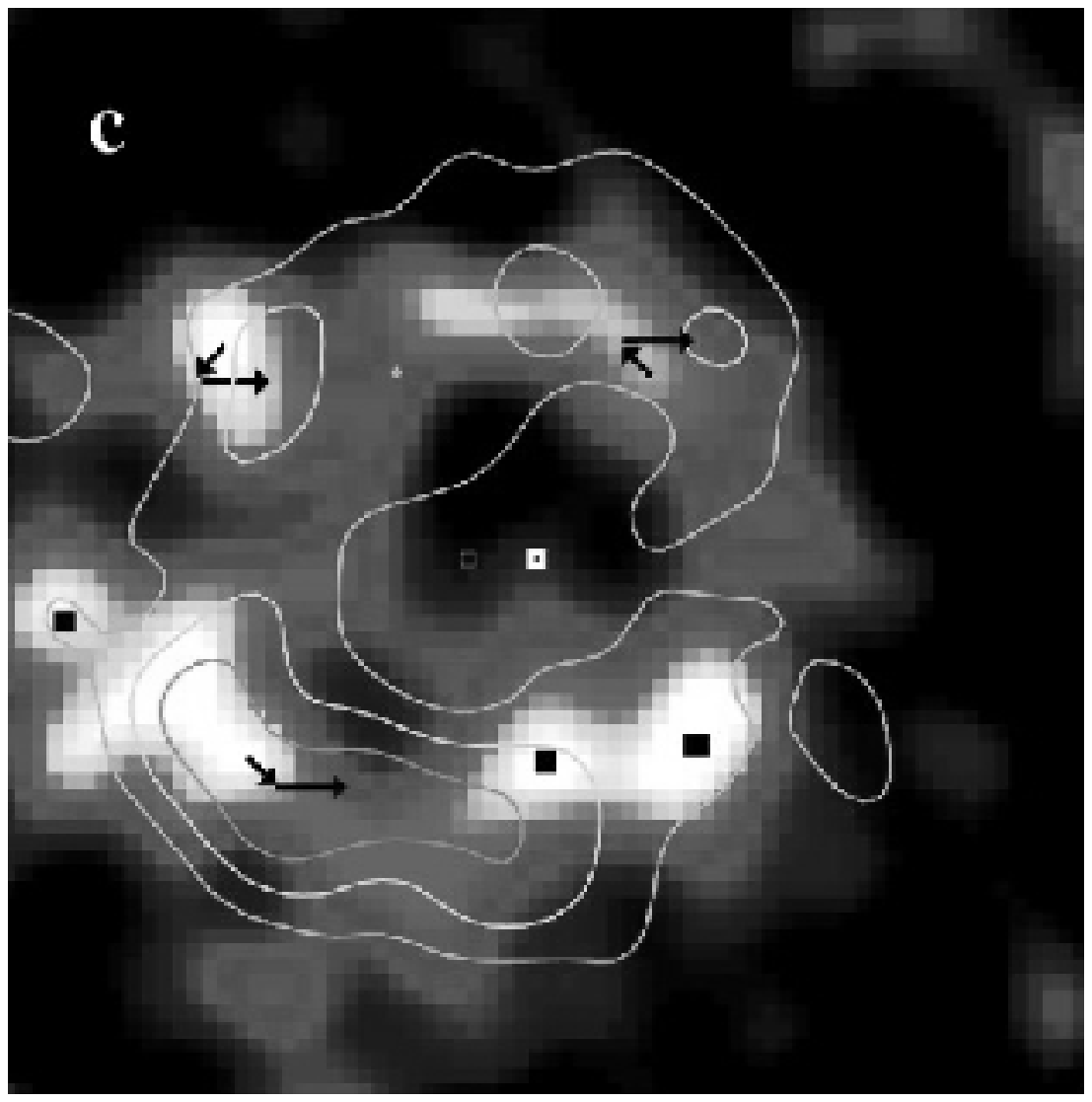}{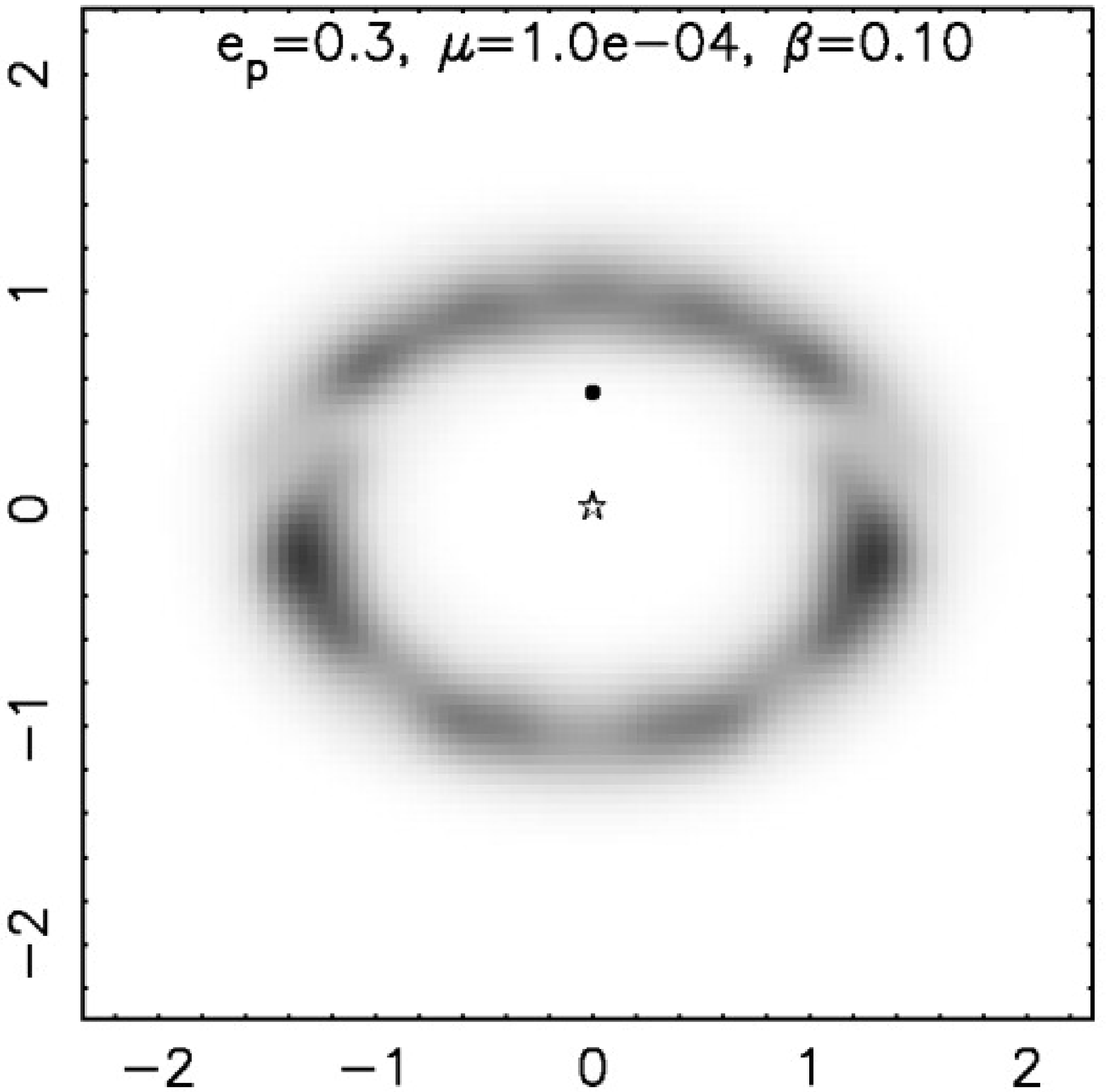}
\caption{(a) The 1997 - 1998 JCMT SCUBA 850 $\mu$m map of $\epsilon$ Eridani
is shown in greyscale while the 2000 - 2002 850 $\mu$m map is shown with
overlaid contours at 30\%, 50\%, and 70\% of the peak surface brightness. Black
squares are suggested background features and black arrows show the orbiting 
motion of the brightness peaks. (Figure taken from Greaves et al. 2005) (b) 
Simulated intensity distribution of the $\epsilon$ Eridani disk, assuming that 
the brightness peaks are generated by dust trapped in the 5:3 and 3:2 exterior 
mean motion resonances of a 30 $M_{\earth}$ planet, with eccentricity, $e$ = 
0.3, and semimajor axis, $a$ = 40 AU. (Figure taken from Quillen \& Thorndike 
2002)}
\end{figure}

If a planet in a debris disk has an eccentric orbit, then the planet may
force circumstellar dust grains into elliptical orbits. Since dust grains at 
pericenter will be closer to the star and therefore warmer than grains at 
apocenter, disks with eccentric planets may possess brightness asymmetries
(Wyatt et al. 1999). High resolution thermal infrared and submillimeter 
imaging of Fomalhaut has revealed a 30\% - 15\% brightness asymmetry in its 
disk ansae (Stapelfeldt et al. 2004; Holland et al. 2003). The dust 
grains in this disk may have experienced secular perturbations of their orbital
elements by a planet, with $a$ = 40 AU and $e$ = 0.15, which forces grains into
an elliptical orbit with the star at one focus (Wyatt et al. 1999; Stapelfeldt 
et al. 2004). Recent \emph{HST} ACS scattered light observations of Fomalhaut,
shown in Figure 3, have confirmed that the star is not at the center of the 
dust grain orbits. Kalas, Graham, \& Clampin (2005) measure an offset of 
$\sim$15 AU between the geometric center of the disk and the position of the 
central star. A 15\% - 5\% brightness asymmetry is observed toward HR 4796A at 
near-infrared and mid-infrared wavelengths (Weinberger, Schneider, \& Becklin 
2000; Telesco et al. 2000), that may be explained if the orbit of HR 4796B has 
an eccentricity $e$ = 0.13 or if there is a $>$0.1 M$_{Jup}$ mass planet at the
inner edge of the disk at 70 AU (Wyatt et al. 1998). 

\begin{figure}
\plottwo{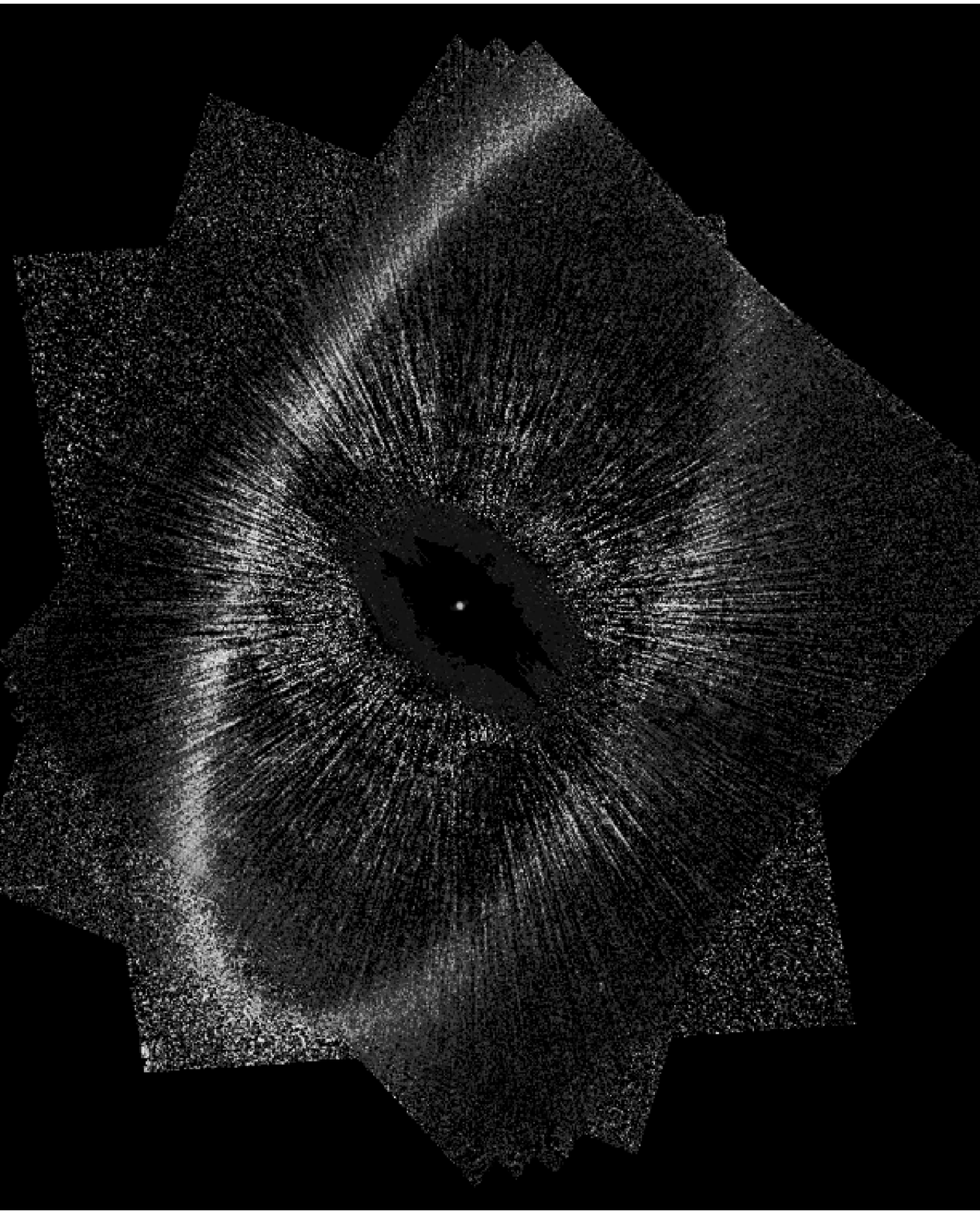}{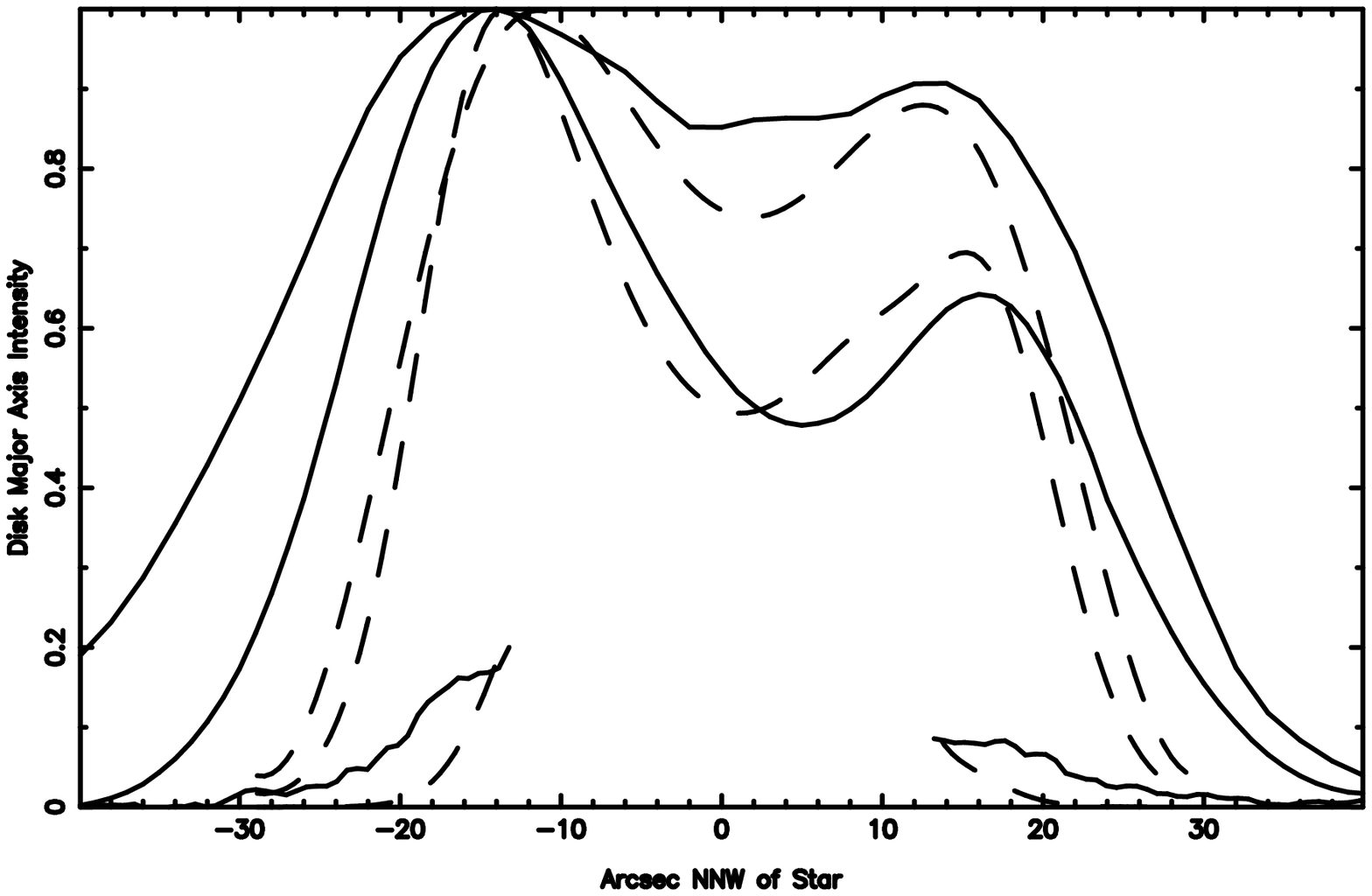}
\caption{(a) ACS scattered light observations of Fomalhaut with North up and
East to the left. The star is offset from the geometric center of the disk by
15 AU, with the inner edge of the disk located at 133 AU - 158 AU from the 
star. (Courtesy P. Kalas, University of California, Berkeley); (b) The solid
lines are line cuts, along the disk axis, through \emph{Spitzer} MIPS 
24 $\mu$m, 70 $\mu$m, and 160 $\mu$m images (from bottom to top) of Fomalhaut. 
The dashed lines are the profiles expected if a planet with $a$ = 40 AU, $e$ = 
0.15 forces the eccentricity of dust grains in the disk. (Figure taken from 
Stapelfeldt et al. 2004)}
\end{figure}

There is tantalizing evidence to suggest that giant planets and debris disks
are correlated. A near-infrared coronagraphic survey of six FGK stars with 
radial velocity planets resolved disks around three objects (Trilling et al. 
2000); however, NICMOS observations were unable to confirm the presence of a 
disk around one of the objects: 55 Cnc (Schneider et al. 2001). \emph{Spitzer} 
MIPS observations of 26 planet-bearing FGK stars have discovered six objects 
that possess 70 $\mu$m excesses, corresponding to a disk fraction 23\% 
(Beichman et al. 2005a), higher than the $\sim$15\% observed toward field 
stars. The MIPS observations suggest that both the frequency and the magnitude 
of dust emission are correlated with the presence of known planets. Since the 
70 $\mu$m excess is generated by cool dust ($T_{gr}$ $<$ 100 K) located beyond 
10 AU, well outside the orbits of the discovered planets, the process which 
correlates the radial velocity planets at $<$5 AU with cool dust beyond 10 AU 
is not understood. By contrast, a submillimeter survey of eight planet-bearing 
stars found no thermal emission toward any of the stars surveyed (Greaves et 
al. 2004). Radial velocity studies of main sequence stars that possess giant 
planets find a correlation between the presence of an orbiting planet and
the metallicity of the central star (Fischer \& Valenti 2005). If correlations 
exist between metallicity and the presence of a planet and between the presence
of a planet and a 70 $\mu$m excess, there might also be a correlation between 
the presence of IRS excess and stellar metallicity. However, no correlations 
between the presence of 70 $\mu$m excess and stellar metallicity have been 
found thusfar (Bryden et al. 2005; Beichman et al. 2005a). 

\section{Infalling Bodies and the Stable Gas Around $\beta$ Pictoris}
Ultra violet and visual spectra of $\beta$ Pictoris possess time-variable, 
high velocity, red-shifted absorption features, initially discovered in 
Ca II H and K and Na I D and later in a suite of metal atoms (including C I, 
C IV, Mg I, Mg II, Al II, Al III, Si II, S I, Cr II, Fe I, Fe II, Ni II, 
Zn II); these features vary on timescales as short as hours and are 
non-periodic (Vidal-Madjar, Lecavelier des Etangs, \& Ferlet 1998). The 
excitation of the atoms, the chemical abundance, the high measured velocities, 
and the rapid time-variability of the gas suggest that the material is 
circumstellar rather than interstellar. The velocity of the atoms, typically 
100 km/sec - 400 km/sec, is close to the free fall velocity at a few stellar 
radii, suggesting that the absorption is produced as stellar photons pass
through the coma of infalling refractory bodies at distances $<$6 AU from the 
star (Karmann, Beust, \& Klinger 2001; Beust et al. 1998). At these distances, 
refractory materials may sublimate from the surface of infalling bodies and 
collisions may produce highly ionized species such as C IV and Al III. If 
infalling bodies generate the observed features, then the fact that the 
features are preferentially red-shifted (rather than equally red- and 
blue-shifted) suggests that some process aligns their orbits. Scattering of 
bodies by a planet on an eccentric orbit (Levison, Duncan, \& Wetherill 1994) 
and bodies in mean motion resonances with an eccentric planet, whose orbits
decay via secular resonant perturbations by the planet (Beust \& Morbidelli 
1996), have been used to explain the preferentially red-shifted features; 
however, both models have difficulties.

The $\beta$ Pictoris disk also possesses a stable component of gas at the 
velocity of the star. Many of the species that possess high velocity features 
also possess very low velocity features that vary slightly over timescales of 
hours to years. Spatially resolved visual spectra of $\beta$ Pic have revealed 
the presence of a rotating disk of atomic gas, observed via emission from Fe I,
Na I, Ca II, Ni I, Ni II, Ti I, Ti II, Cr I, and Cr II. Estimates of the 
radiation pressure acting on Fe and Na atoms suggest that these species should 
be accelerated to terminal velocities $\sim$100s - 1000s km/sec, significantly 
higher than is observed (Brandeker et al. 2004). One possible explanation for 
the low velocity of the atomic gas is that the gas is ionic and that Coulomb 
interactions between ions reduce the effective radiation pressure on the bulk 
gas. Fern\'{a}ndez, Brandeker, \& Wu (2005) suggest that ions in the disk 
couple together into a fluid, with an effective radiation pressure coefficient,
that is bound to the system and that brakes the gas if $\beta_{eff}$ $<$ 0.5. 
In particular, they suggest that atomic carbon may be important for reducing 
the effective radiation pressure coefficient if the carbon abundance is 
$>$10$\times$ solar because the expected ionization fraction of atomic carbon 
is 0.5 and $F_{rad}/F_{grav}$ $\approx$ 0. Measurements of the line-of-sight 
abundance of ionized carbon, inferred from C II absorption in the O VI 
$\lambda$1038 emission using \emph{FUSE}, confirm that the ionization fraction
of carbon is $\sim$0.5 and suggest that carbon is overabundant by 
approximately a factor of 10 compared with measurements of the stable gas from 
the literature (Roberge et al. 2006, in preparation). Estimates of the total 
gas mass, inferred from measured elemental abundances and the gas density 
radial profile from scattered emission, suggest that the $\beta$ Pic disk 
contains $\sim$0.004 M$_{\earth}$ gas or a gas:dust ratio $\sim$0.1 (Roberge et
al. 2006).

\emph{HST} GHRS and STIS observations have also detected stable CO and C I 
($^{3}$P) absorption at the velocity of the star (Jolly et al. 1998; Roberge 
et al. 2000). Since CO and C I are expected to be photodissociated and 
photoionized by interstellar UV photons on timescales $\sim$200 years, these 
gases must be replenished from a reservoir. One possibility is that the CO is 
produced by slow sublimation of orbiting comets at serveral 10s of AU from the 
star; however, the observed CO possesses a low $^{12}$CO:$^{13}$CO ratio 
(R = 15$\pm$2) compared to solar system comets (R = 89). The overabundance of
$^{13}$CO may be explained if it is produced from $^{12}$CO in the reaction 
\begin{equation}
\textup{$^{13}$C}^{+} + \textup{$^{12}$CO} \rightleftharpoons 
\textup{$^{13}$CO} + \textup{$^{12}$C}^{+} + \textup{35 \ K}
\end{equation} 
at temperatures below 35 K (Jolly et al. 1998). The order of magnitude 
difference in the measured column densities of C I (N(C I) = 
(2-4)$\times$10$^{16}$ cm$^{-2}$; Roberge et al. 2000) and CO (N(CO) = 
2.5$\pm$0.5$\times$10$^{15}$ cm$^{-2}$; K. H. Hinkel, private communication) 
suggest that C I is not produced by photodissociation of CO. Similarly, the 
disparity in the measured excitation temperatures for CO and C I, $T_{ex}$ = 
20-25 K and 80 K, suggests that the observed CO and C I are not cospatial 
(Roberge et al. 2000). One possibility is that the observed stable 
C I ($^{3}$P) is produced directly by sublimation of infalling bodies; if there
are 100 bodies per year, then Roberge et al. (2000) estimate that each 
infalling object generates a C I ($^{3}$P) column density, $N_{comet}$ $\sim$ 
10$^{11}$ cm$^{-2}$ if C I is only destroyed via ionization by interstellar 
photons with $\Gamma$ = 0.004 year$^{-1}$, consistent with evaporation of 
kilometer-sized objects.

\section{Future Work}
The age $\sim$10 Myr appears to mark a transition in the gross properties of 
circumstellar disks. Pre-main sequence objects, such as T-Tauri and Herbig 
Ae/Be stars, possess optically thick, gaseous disks while young, main
sequence stars possess optically thin, gas-poor disks. Recent studies of the 
10 Myr old TW Hydrae association (Weinberger et al. 2004) and the 30 Myr 
Tucana-Horologium association (Mamajek et al. 2004) find that warm 
circumstellar dust ($T_{gr}$ = 200 - 300 K), if present around young stars
in these associations, reradiates less than 0.1\% - 0.7\% of the stellar 
luminosity, $L_{IR}/L{*}$ $<$ (1 - 7)$\times$10$^{-3}$. If the majority of 
stars possess planetary systems at an age of $\sim$10 Myr, why do some stars, 
such as TW Hydrae, Hen 3-600, HD 98800, and HR 4796A in the TW Hydrae 
association, still possess large quantities of gas and/or dust? To investigate 
the origin of the dispersion in disk properties around $\sim$10 Myr old stars, 
I have begun a multi-wavelength study of solar-like stars in the 5 - 20 Myr old
(Preibisch et al. 2002; Mamajek et al. 2002) Scorpius-Centaurus OB Association,
located at a distance of 118 - 145 pc away from the Sun.

One component of my study is a \emph{Spitzer} MIPS 24 $\mu$m and 70 $\mu$m 
search for infrared excess around $\sim$100 F- and G-type stars in Sco-Cen;
the first results suggest that fractional 24 $\mu$m excess luminosity and 
fractional x-ray luminosity may be anti-correlated (Chen et al. 2005a). 
Observations of the first 40 targets detect strong 24 $\mu$m and 70 $\mu$m 
excess around six stars, corresponding to $L_{IR}/L_{*}$ = 7$\times$10$^{-4}$
- 1.5$\times$10$^{-2}$, and weak 24 $\mu$m excess around seven others, 
corresponding to $L_{IR}/L_{*}$ = 3.5$\times$10$^{-5}$ - 4.4$\times$10$^{-4}$.
Only one of the 24 $\mu$m and 70 $\mu$m bright disks is detected by 
\emph{ROSAT} while four of the weak 24 $\mu$m excess sources are. The presence 
of strong stellar winds, $\sim$1000$\times$ larger than our solar wind, may 
explain the depletion of dust in disks around x-ray emitting sources. Follow-up
\emph{Spitzer} IRS 5 - 35 $\mu$m spectroscopy may allow us to determine 
whether corpuscular stellar wind effectively removes dust grain in x-ray 
emitting systems. Since the inward drift velocity of dust grains under 
corpuscular stellar wind and Poynting-Robertson drag is 
(1 + ${\dot M_{wind}} c^{2}/L_{*}$) larger than under Poynting-Robertson drag 
alone, a disk influenced by strong stellar wind is expected to possess a 
constant surface density distribution. If the grains are large, then the 
infrared spectra will be well modeled by $F_{\nu}$ $\propto$ $\lambda$. 
However, if collisions and ejection by radiation pressure is the dominant grain
removal mechanism, then the infrared spectra may be better modeled by a black 
body. 

Other components of my study include \emph{Spitzer} IRS 5 - 35 $\mu$m 
spectroscopy and MIPS SED mode observations of MIPS-discovered excess sources 
to search for emission from silicates and water ice and to constrain the 
composition and grain size. The minimum grain size, due to blow-out from 
radiation pressure, around stars in this sample is $a_{min}$ $\sim$0.5 - 2.0 
$\mu$m, small enough to produce spectral features. Ground-based 10 $\mu$m 
spectroscopy of HD 113766, an F3 binary member of the Sco-Cen subgroup Lower 
Centaurus Crux, suggests that the dust around this object is highly processed 
by an age of $\sim$16 Myr. Fits to the 10 $\mu$m silicate feature suggest that 
$>$90\% of the amorphous silicate mass is contained in large grains (with radii
$a$ $\sim$ 2 $\mu$m) and that $\sim$30\% of the mass is contained in 
crystalline forsterite (Schutz, Meeus, \& Sterzik 2005).

%Measuring the timescale for gas dissipation in circumstellar disks is crucial
%for constraining the mechanism by which giant planets form. High resolution
%(R = 100,000), mid-infrared spectroscopy using TEXES will be critical for 
%detecting narrow emission features expected from circumstellar gas. H$_{2}$
%S(2) emission at 12.28 $\mu$m is detected toward AB Aur, an isolated Herbig Ae
%star, using TEXES on the IRTF (M. Bitner, private communication). Atomic gases
%may be good tracers for H$_{2}$. Self-consistent models for the dust and gas 
%in optically thin disks around solar-like stars suggest that [S I] 25.23 
%$\mu$m emission may be straight-forward to detect with \emph{Spitzer} IRS 
%(R $\sim$ 600). If the gas has solar sulfur abundance and all of the sufur is 
%in the gas phase (not incorporated into grains), then the line:continuum ratio
%may be as high as 3.5 for a disk with 10$^{-2}$ $M_{Jup}$ gas and 
%10$^{-5}$ $M_{Jup}$ dust compared to a line:continuum ratio of $<$5\% for 
%H$_{2}$ S(0) and S(1) emission at 28.2 $\mu$m and 17.0 $\mu$m (Gorti \& 
%Hollenbach 2004). 

\acknowledgements I would like to thank the Astronomy Department at the 
University of Texas at Austin for giving me the opportunity to write this 
review and M. Jura, P. Kalas, W. Liu, J. Najita, and A. Roberge for stimulating
conversations and correspondence during the preparation of this manuscript. 
Support for this work was provided by the NASA through the Spitzer Fellowship 
Program under award NAS7-1407.

%%% THE BIBLIOGRAPHY
%%%
%%% CONSULT SECTION 3 OF "INSTRUCTIONS FOR AUTHORS" FOR HOW TO USE NATBIB.
%%% AUTHORS ARE ENCOURAGED TO USE EITHER THE "THEBIBLIOGRAPY" ENVIRONMENT
%%% BY UNCOMMENTING (DELETING THE "%" SYMBOL) THE COMMANDS BELOW, OR BY
%%% USING THE BIBTEX ENVIRONMENT. TO FIND OUT WHICH IS APPLICABLE TO YOUR
%%% CONTRIBUTION, CONSULT THE VOLUME EDITORS FOR YOUR PROCEEDINGS.
%%%

\end{document}